\documentclass[twocolumn,english,prl,floatfix,citeautoscript,nofootinbib]{revtex4}
\usepackage{amsbsy}
\usepackage{latexsym,epsfig,graphicx}
\usepackage{dcolumn}
\usepackage{subfigure}
\usepackage{comment}
\usepackage{color}
\usepackage[colorlinks,urlcolor=blue,citecolor=blue]{hyperref}
\usepackage{amstext}
\usepackage{amssymb}
\usepackage{setspace}

\begin{document}

\title{Spin-tensor--momentum-coupled Bose-Einstein condensates}
\author{Xi-Wang Luo}
\author{Kuei Sun}
\author{Chuanwei Zhang}
\thanks{Corresponding author. \\
Email: \href{mailto:chuanwei.zhang@utdallas.edu}{chuanwei.zhang@utdallas.edu}%
}
\affiliation{Department of Physics, The University of Texas at Dallas, Richardson, Texas
75080-3021, USA}

\begin{abstract}
The recent experimental realization of spin-orbit coupling for ultracold
atomic gases provides a powerful platform for exploring many interesting
quantum phenomena. In these studies, spin represents spin vector (spin-1/2
or spin-1) and orbit represents linear momentum. Here we propose a scheme to
realize a new type of spin-tensor--momentum coupling (STMC) in spin-1
ultracold atomic gases. We study the ground state properties of interacting
Bose-Einstein condensates (BECs) with STMC and find interesting new types of
stripe superfluid phases and multicritical points for phase transitions.
Furthermore, STMC makes it possible to study quantum states with dynamical
stripe orders that display density modulation with a long tunable period and
high visibility, paving the way for direct experimental observation of a new
dynamical supersolid-like state.. Our scheme for generating STMC can be
generalized to other systems and may open the door for exploring novel
quantum physics and device applications.
\end{abstract}

\maketitle

\emph{Introduction}.---The coupling between matter and gauge field plays a
crucial role for many fundamental quantum phenomena and practical device
applications in condensed matter \cite{xiao2010berry, hasan2010colloquium,
qi2011topological} and atomic physics \cite{dalibard2011colloquium}. A
prominent example is the spin-orbit coupling, the coupling between a
particle's spin and orbit (e.g., momentum) degrees of freedom, which is
responsible for important physics such as topological insulators and
superconductors \cite{hasan2010colloquium, qi2011topological}. In this
context, recent experimental realization of spin-orbit coupling in ultracold
atomic gases \cite{lin2011spin, zhang2012collective, qu2013observation,
olson2014tunable, wang2012spin, cheuk2012spin, Williams2013,
huang2016experimental, wu2016realization} opens a completely new avenue for
investigating quantum many-body physics under gauge field \cite%
{Stanescu2008, zhang2008p, Wu2011, wang2010spin, ho2011bose, li2012quantum,
zhang2012mean, hu2012spin, ozawa2012stability, Gong2011, Hu2011, Yu2011,
Qu2013b, Zhang2013b, galitski2013spin}.

So far in most works on spin-orbit coupling in solid state and cold atomic
systems, the spin degrees of freedom are taken as rank-1 spin vectors $F_{i}$
($i=x,y,z$), such as electron spin-1/2 or pseudospins formed by atomic
hyperfine states that can be large (e.g., spin-1 or 3/2). Experimentally,
spin-orbit coupling for spin-1 Bose-Einstein condensates (BECs) has been
realized recently \cite{campbell2015itinerant, luo2016tunable} and
interesting magnetism physics has been observed \cite{lan2014raman,
Natu2015,sun2016interacting, yu2016phase, martone2016tricriticalities}.
Mathematically, it is well known that there exist not only spin vectors, but
also spin tensors [e.g., irreducible rank-2 spin-quadrupole tensor $%
N_{ij}=\left( F_{i}F_{j}+F_{j}F_{i}\right) /2-\delta_{ij}\mathbf{F}^2/3$] in
a large spin ($\geq 1$) system. Therefore two natural questions are: \textit{%
i}) Can the coupling between spin tensors of particles and their linear
momenta be realized in experiments? \textit{ii}) What new physics may emerge
from such spin-tensor--momentum coupling (STMC)?

In this Letter, we address these two questions by proposing a simple
experimental scheme for realizing STMC for spin-1 ultracold atomic gases.
Our scheme is based on slight modification of previous experimental setup
\cite{campbell2015itinerant} and is experimentally feasible. The STMC
changes the band structure dramatically, leading to interesting new physics
in the presence of many-body interactions between atoms. Although both
bosons and fermions can be studied, here we only consider spin-1 BECs to
illustrate the effects of STMC. Our main results are:

\textit{i}) The single particle band structure with STMC consists of two
bright-state bands (top and bottom) and one dark-state middle band [Fig.~\ref%
{fig:sys}(b)], where the dark-state band is not coupled with two
bright-state bands through Raman coupling. However, the dark-state band
plays an important role on both ground-state and dynamical properties of the
interacting BECs.

\begin{figure}[t]
\includegraphics[width=1.0\linewidth]{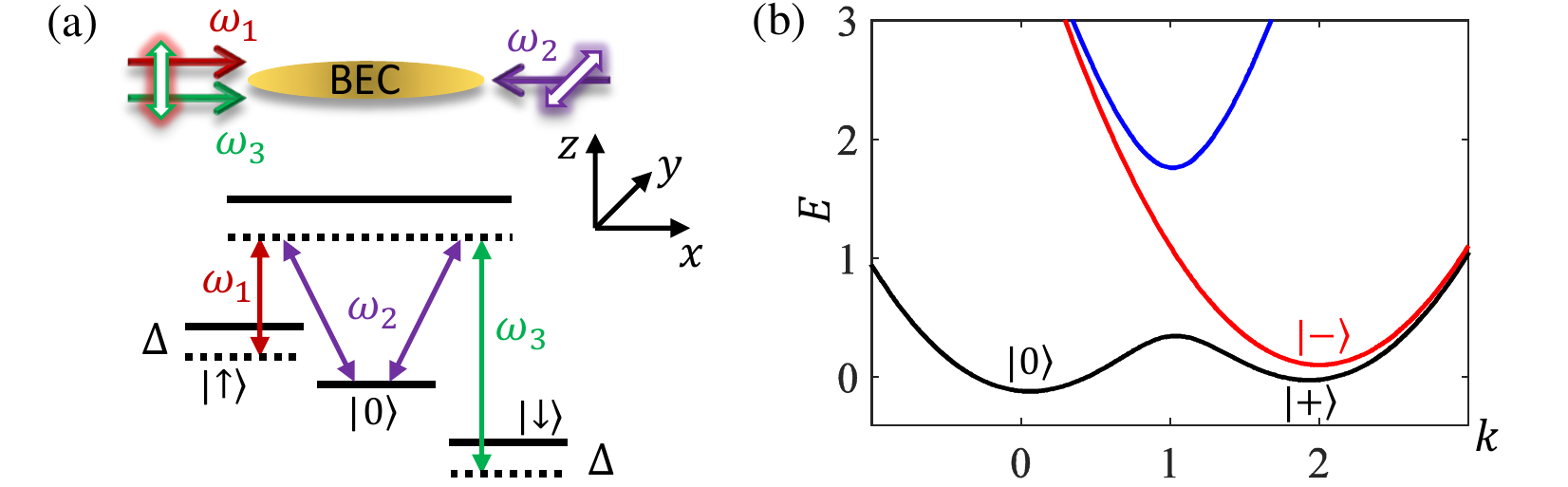}
\caption{(a) Top: Experimental scheme to generate STMC in BEC. Bottom: Raman
transitions between three hyperfine spin states with detuning $\Delta $. (b)
Single-particle band structure for Raman strength $\Omega =0.5$ and detuning
$\Delta =0.1$. The (dominant) spin components $\left\vert 0\right\rangle $
and $\left\vert \pm \right\rangle =\frac{1}{\protect\sqrt{2}}(\left\vert
\uparrow \right\rangle \pm \left\vert \downarrow \right\rangle )$ are
indicated around the corresponding band minima.}
\label{fig:sys}
\end{figure}

\textit{ii}) We study the ground-state phase diagrams with exotic plane-wave
and stripe phases, where the dark-state middle band can be partially
populated despite not the single particle ground state. The stripe phase is
a coherent superposition of two or more plane-wave states. It possesses both
superfluid property as a BEC and crystal-like density modulation that
spontaneously breaks translational symmetry of the Hamiltonian, satisfying
two major criteria for the supersolid order~\cite{pomeau1994dynamics}.
Experimentally, the stripe order has recently been observed indirectly using
Bragg reflection \cite{li2017stripe}. We find the transitions between
different phases possess interesting multicriticality phenomena with triple,
quadruple and even quintuple points.

\textit{iii}) The existence of dark middle band makes it possible to study
quantum states with \textit{dynamical supersolid-like stripe orders}. In
particular, we show how to dynamically generate a stripe state with a long
tunable period ($\sim 5\mu $m) and high visibility ($\sim 100\%$) of density
modulation, which may be directly measured in experiments (such direct
measurement is still challenging for the ground-state stripe patterns due to
their short period and low visibility \cite{li2014superstripes}). The
dynamical stripe state as a superfluid BEC, although not the ground state,
does possess interesting stripe patterns that break the translational
symmetry of the Hamiltonian, resembling a dynamical supersolid-like order.

\emph{The model}.---We consider a setup similar as that in the recent
experiment \cite{campbell2015itinerant} but with a slightly different laser
configuration, as shown in Fig.~\ref{fig:sys}(a), where three Raman lasers
with wavenumber $k_{\text{R}}$ are employed to generate STMC. The three
lasers induce two Raman transitions between hyperfine spin states $|0\rangle
$ and $\left\vert \uparrow (\downarrow )\right\rangle $, both of which have
the same recoil momentum $2k_{\text{R}}$ along the $x$ direction. The
single-particle Hamiltonian in the spin-1 basis $(\left\vert \uparrow
\right\rangle ,|0\rangle ,\left\vert \downarrow \right\rangle )^{T}$ is (we
set $\hbar=1$)
\begin{equation}
\widetilde{H}_{0}=-\frac{{\mathbf{\nabla }}^{2}}{2m}+\Delta F_{z}^{2}+\left(
\sqrt{2}\Omega e^{i2k_{\text{R}}x}|0\rangle \langle +|+h.c.\right) ,
\end{equation}%
where $F_{z}^{2}=\left\vert \uparrow \right\rangle \left\langle \uparrow
\right\vert +\left\vert \downarrow \right\rangle \left\langle \downarrow
\right\vert $ is equivalent to the spin tensor $N_{zz}$ (up to a constant), $%
|+\rangle \equiv \frac{1}{\sqrt{2}}(\left\vert \uparrow \right\rangle
+\left\vert \downarrow \right\rangle )$, $\Omega $ is the Raman coupling
strength, and $\Delta $ is the detuning for both $\left\vert \uparrow
\right\rangle $ and $\left\vert \downarrow \right\rangle $ states. We see
that another spin state $|-\rangle \equiv \frac{1}{\sqrt{2}}(\left\vert
\uparrow \right\rangle -\left\vert \downarrow \right\rangle )$ is always an
eigenstate and does not couple to $|0\rangle $ nor $|+\rangle $ through $%
\Omega $, and thus is a dark state.

Since the BEC wavefunction in the $y$ and $z$ directions is not affected by
the Raman lasers, we can consider the physics only along the $x$ direction~%
\cite{sun2016interacting, yu2016phase, martone2016tricriticalities}. After a
unitary transformation $U=\exp (-i2k_{\text{R}}xF_{z}^{2})$ to
quasi-momentum basis, we write the Hamiltonian in energy and momentum units $%
\frac{k_{R}^{2}}{2m}$ and $k_{R}$, respectively, as
\begin{equation}
H_{0}=-\partial _{x}^{2}+(\Delta +4+4i\partial _{x})F_{z}^{2}+\sqrt{2}\Omega
F_{x},  \label{eq:H0}
\end{equation}%
where $\Omega $ and $\Delta $ are dimensionless transverse-Zeeman and
spin-tensor potential, respectively, and $(i\partial _{x})F_{z}^{2}$
describes the coupling between spin tensor $F_{z}^{2}$ and the linear
momentum, \textit{i.e.}, STMC.

The single-particle Hamiltonian has three energy bands [see a typical
structure in Fig.~\ref{fig:sys}(b)]. The dark-state middle band always has
the spin state $|-\rangle $ and spectrum $(k-2)^{2}+\Delta $, which are
independent of $\Omega $. The top and bottom bright-state bands exhibit the
same behavior as the known spin-orbit-coupled spin-$1/2$ system with spin
states $|0\rangle $ and $|+\rangle $. The decoupling of the middle band is
protected by the spin-tensor symmetry $[F_{x}^{2},H_{0}]=0$, under which the
middle band (top and bottom bands) corresponds to $\langle F_{x}^{2}\rangle
=0$ (1). Although the single-particle ground state always selects the bottom
band, the atomic interactions can break the symmetry and drastically change
the BEC's ground state as well as dynamical properties by involving the
middle band.

Under the Gross-Pitaevskii (GP) mean-field approximation, the energy density
becomes
\begin{equation}
\varepsilon =\frac{1}{V}\int dx\left[ \Psi ^{\dag }H_{0}\Psi +\frac{g_{0}}{2}%
(\Psi ^{\dag }\Psi )^{2}+\frac{g_{2}}{2}(\Psi ^{\dag }\mathbf{F}_{U}\Psi
)^{2}\right] ,  \label{eq:energydensity}
\end{equation}%
with $V$ the system volume, and $\Psi $ the three-component condensate
wavefunction normalized by the average particle number density $\bar{n}%
=V^{-1}\int dx\Psi ^{\dag }\Psi $. The interaction strengths $g_{0,2}$
represent density and spin interactions in spinor condensates~\cite%
{ho1998spinor, Ohmi1998Bose}, respectively. $\mathbf{F}_{U}=U^{\dag }\mathbf{%
F}U$ is the unitarily transformed spin operator, whose $x$ and $y$
components exhibit spatial modulation that cannot be eliminated through any
local spin rotation (different from previous models~\cite%
{sun2016interacting, yu2016phase, martone2016tricriticalities}). Such
modulation is essential for stripe phases in the system.

We consider a variational ansatz \cite{SM}%
\begin{equation}
\Psi =\sqrt{\bar{n}}\left( |c_{1}|\chi _{1}e^{ik_{1}x}+|c_{2}|\chi
_{2}e^{ik_{2}x+i\alpha }\right)  \label{eq:ans}
\end{equation}%
to find the ground state, with $|c_{1}|^{2}+|c_{2}|^{2}=1$, and spinors $%
\chi _{j}=(\cos \theta _{j}\cos \phi _{j},-\sin \theta _{j},\cos \theta
_{j}\sin \phi _{j})^{T}$. The energy density now becomes a functional of
eight variational parameters $|c_{1}|$, $k_{1}$, $k_{2}$, $\theta _{1}$, $%
\theta _{2}$, $\phi _{1}$, $\phi _{2}$, and $\alpha $, and its minimization (%
$\varepsilon_{\text{g}}=\min\{\varepsilon\}$) leads to the ground state~\cite%
{SM}. The quantum phase diagram can be characterized by the variational
wavefunction, experimental observables $\langle F_{z}\rangle $ and $\langle
F_{z}^{2}\rangle $, and the symmetry $\langle F_{x}^{2}\rangle $. The
derivative of the ground-state energy $\frac{\partial \varepsilon_{\text{g}}%
}{\partial \Delta}= \langle F^2_z\rangle$ ($\frac{\partial^2 \varepsilon_{%
\text{g}}}{\partial \Delta^2}= \frac{\partial \langle F^2_z\rangle}{\partial
\Delta}$) displays discontinuity as $\Delta$ varies across a first-order
(second-order) phase boundary~\cite{SM}. This argument also applies to $%
\frac{\partial \varepsilon_{\text{g}}}{\partial \Omega}$ ($\frac{\partial^2
\varepsilon_{\text{g}}}{\partial \Omega^2}$)~\cite{SM}. We also numerically
solve the GP equation using imaginary time evolution to obtain the ground
states, which are in good agreement with the variational results.

\begin{figure}[tbp]
\includegraphics[width=1.0\linewidth]{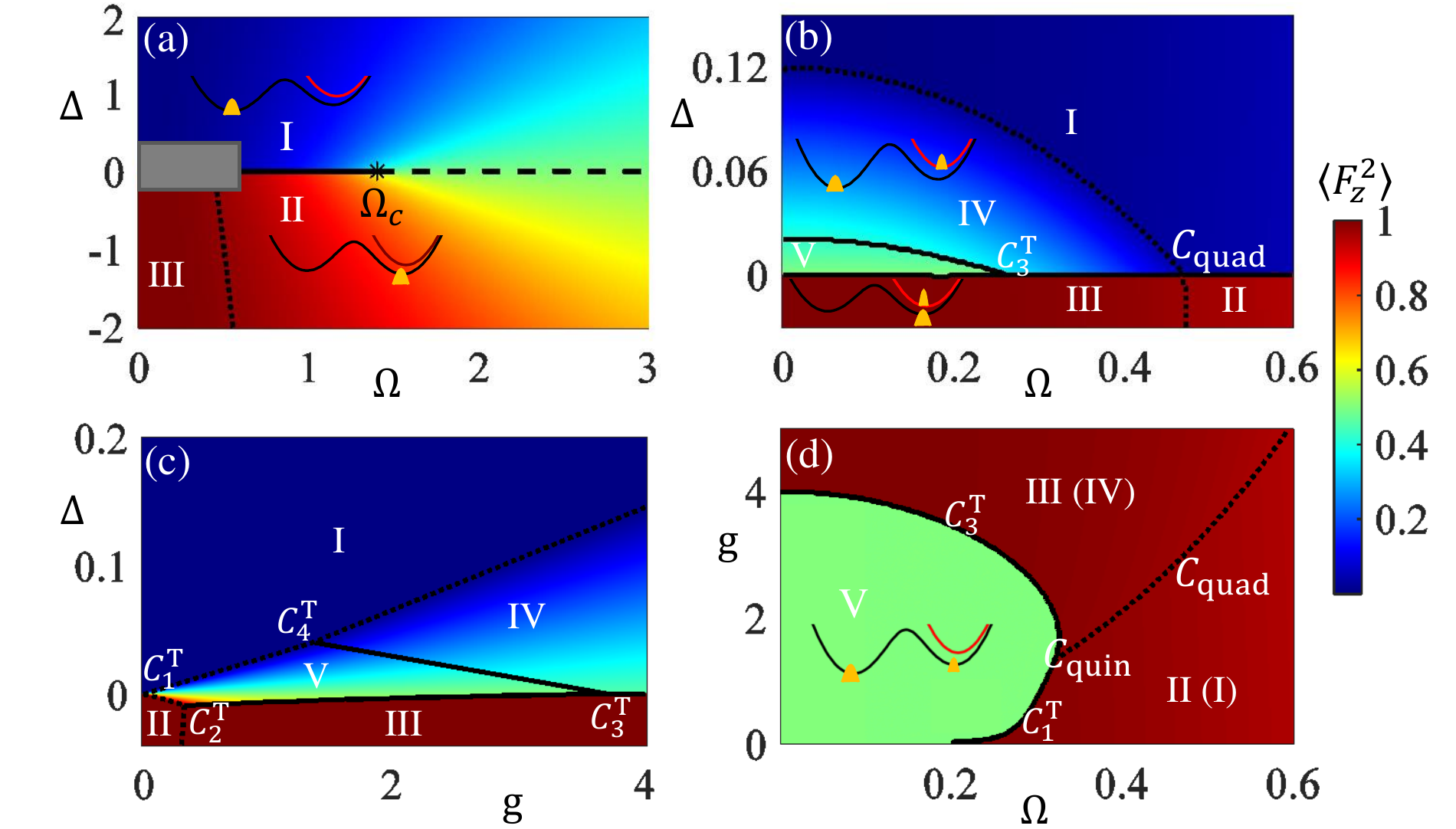}
\caption{ (a) Ground state phase diagram in the $\Omega$-$\Delta$ plane with
$g\bar{n}=3$. The dashed line is a crossover boundary. (b) Zoom in of the
framed region in (a). (c) [(d)] Ground state phase diagram in the $g$-$%
\Delta $ ($g$-$\Omega$) plane with $\Omega=0.16$ ($\Delta=0$). Solid
(dotted) lines represent first (second) order phase transitions. The
interaction ratio is $g_0=-50g_2\equiv g$.}
\label{fig:phase}
\end{figure}

\emph{Phase diagram}.---For ferromagnetic interaction $g_{2}<0$ (e.g., $^{87}$%
Rb), the BEC has three plane-wave ($|c_{1}c_{2}|=0$) and two stripe ($%
|c_{1}c_{2}|\neq 0$) phases (Fig.~\ref{fig:phase}): (I) plane-wave phase in $%
k<1$, having $\langle F_{z}\rangle =0$ (spin unpolarized), $\langle
F_{z}^{2}\rangle <0.5$, and $\langle F_{x}^{2}\rangle =1$ (middle band
unpopulated); (II) plane-wave phase in $k>1$, having $\langle F_{z}\rangle
=0 $, $\langle F_{z}^{2}\rangle >0.5$, and $\langle F_{x}^{2}\rangle =1$;
(III) spin-polarized plane-wave phase in $k>1$ having $\langle F_{z}\rangle
\neq 0$ and $\langle F_{x}^{2}\rangle <1$ (middle band populated); (IV)
mix-band stripe phase, having $k_{1}<1$, $k_{2}>1$, and $\langle
F_{x}^{2}\rangle <1$; (V) bottom-band stripe phase, same as (IV) except $%
\langle F_{x}^{2}\rangle =1$. The last three phases exhibit $Z_{2}$
ferromagnetism: phases (III), (IV), and (V) all have twofold degenerate
ground states with global ferromagnetic order $\pm \langle F_{z}\rangle \neq
0$, $\pm \langle F_{y}\rangle \neq 0$, and $\pm \langle F_{x}\rangle \neq 0$%
, respectively. Note that these orders are calculated in the laboratory
frame (the basis of ${\widetilde{H}}_{0}$) and reflect the energetic favor
by the ferromagnetic interaction. For anti-ferromagnetic interaction $%
g_{2}>0 $ (e.g., $^{23}$Na), the system has a relatively simple phase
diagram containing only two plane-wave phases (I) and (II), separated by a
first-order phase-boundary at $\Delta =0$. Hereafter we focus on the
ferromagnetic case.

In Fig.~\ref{fig:phase}(a) we plot the phase diagram in the $\Omega $-$%
\Delta $ plane. At a sufficiently large $\Omega $, the middle band does not
participate in the ground state, so the phase diagram is similar to the
spin-orbit-coupled spin-$1/2$ system: the two plane-wave phases (I) and (II)
are separated by a first-order-transition boundary (solid line along $\Delta
=0$) if $\Omega <\Omega _{c}$ or a crossover one (dashed line) if $\Omega
>\Omega _{c}$. As $\Omega $ decreases, the middle band minimum gets closer
to the right minimum of the bottom band [Fig.~\ref{fig:sys}(b)]. If the BEC
originally stays in the plane-wave phase (II) ($\Delta <0$), it starts to
partially occupy the middle band [Fig.~\ref{fig:phase}(b), bottom inset],
undergoing a second-order transition (dotted curve) to the polarized phase
(III). From the energetic point of view, the BEC populates to a slightly
higher single particle energy state to get polarized to reduce ferromagnetic
interaction energy. Note that phase (III) is still a plane-wave phase since
the BEC occupies both bands at the same $k$.

\begin{figure}[tbp]
\includegraphics[width=1.0\linewidth]{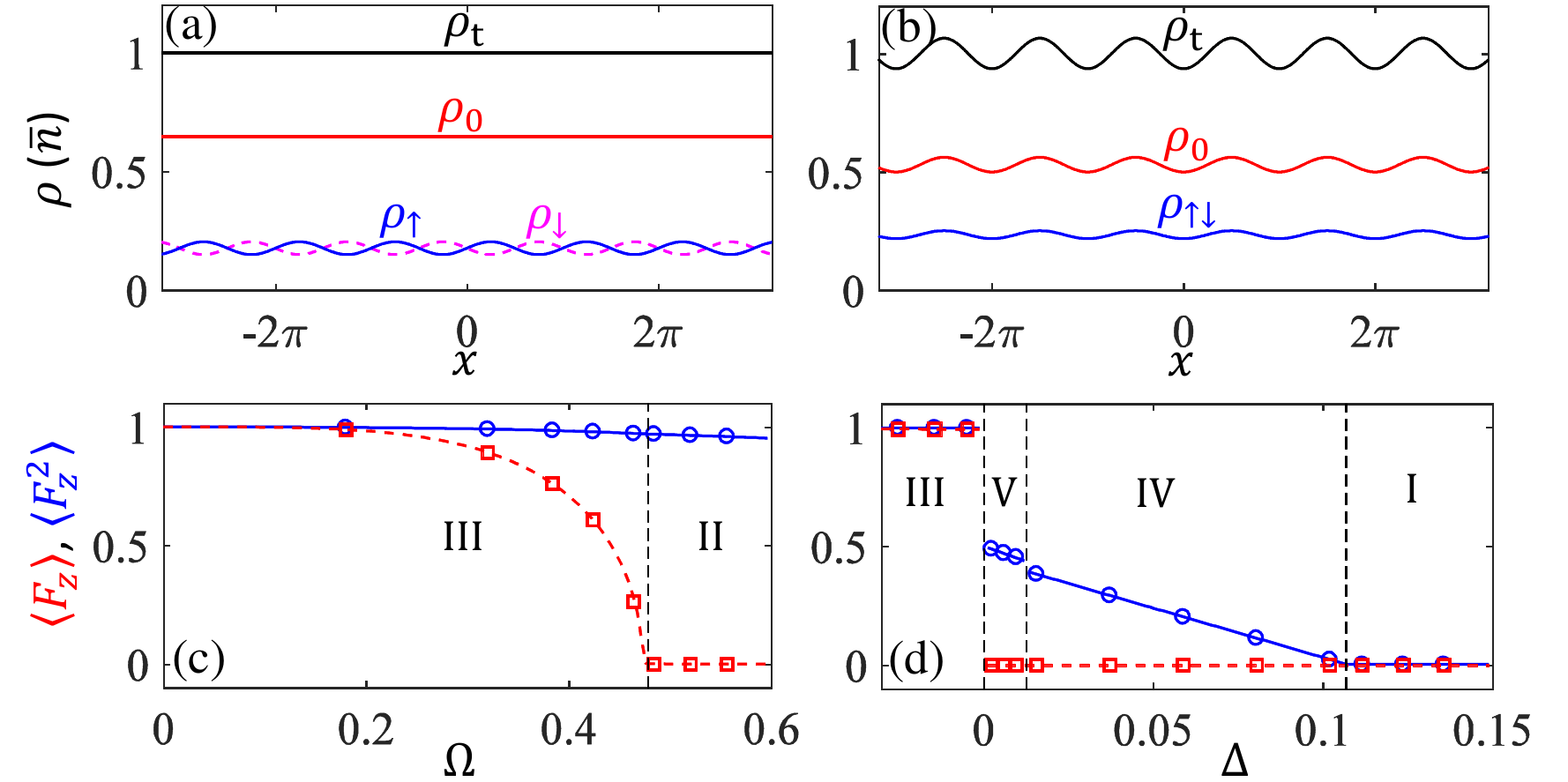}
\caption{(a) [(b)] The local density modulations of phase (IV) [(V)] in Fig.~%
\protect\ref{fig:phase}(b), with $\Omega=0.16$ and $\Delta=0.006$ ($%
\Delta=0.023$). (c) [(d)] $\langle F_z^2\rangle$ (blue-solid line) and $%
\langle F_z\rangle$ (red-dashed line) vs $\Omega$ ($\Delta$) along the path $%
\Delta=-0.018$ ($\Omega=0.16$) in Fig.~\protect\ref{fig:phase}(b). Dots
(lines) are obtained from imaginary time GP equation (variational method). }
\label{fig:GP}
\end{figure}

At a small $\Omega $ and $\Delta >0$, the energy difference between the
single-particle band minimum [plane wave (I)] and the other bottom-band
minimum [plane wave (II)] or the middle-band minimum is comparable to the
interaction energy, so the BEC may favor the co-occupation of (I) and a
higher-energy local minimum as long as the total energy can be reduced more
by the interaction. In Fig.~\ref{fig:phase}(b), we zoom in the framed region
of Fig.~\ref{fig:phase}(a) and show the emergence of two stripe phases. The
mix-band stripe phase (IV) is the superposition of plane wave (I) and the
one around the middle-band minimum (top inset). Phase (IV) exhibits
spin-density waves due to the superposition [Fig.~\ref{fig:GP}(a)] and a
global ferromagnetic order $\langle F_{y}\rangle \neq 0$ that reduces the $%
g_{2}$ interaction energy, compensating the higher middle-band energy. Note
that phase (IV) has a uniform total density due to the orthogonality between
the middle and bottom band spins, but the spin-density waves form a stripe
pattern. The bottom-band stripe phase (V), which appears at even weaker $%
\Omega $ and $\Delta $, is the superposition of two bottom-band plane waves
(I) and (III) [Fig.~\ref{fig:phase}(d) inset]. Phase (V) exhibits a
total-density wave [Fig.~\ref{fig:GP}(b)], which, compared with (IV),
increases the $g_{0}$ interaction energy, but the total energy is favorable
due to the pure bottom-band occupation and global ferromagnetic order $%
\langle F_{x}\rangle \neq 0$. We remark that the superposition of three
plane waves (with co-occupation of three band minima) is never energetically
favorable because it cannot maximize the ferromagnetic order.

Returning to the phase diagram Fig.~\ref{fig:phase}(b), the (I)--(IV) phase
boundary corresponds to a second-order transition, which meets the
(II)--(III) boundary at a quadruple point $C_{\mathrm{quad}}$ at $\Delta =0$%
. The (IV)--(V) boundary corresponds to a first-order transition, which
encounters phase (III) at a triple point $C_{3}^{\mathrm{T}}$ at $\Delta =0$%
. To study the dependence on interaction, we plot the phase diagram in the $%
\Delta $-$g$ plane in Fig.~\ref{fig:phase}(c), with a fixed ratio $%
g_{0}=-50g_{2}\equiv g$. We see that the stripe region increases with $g$
(due to the increasing $g_{2}$), and phase (IV) is more favorable than (V)
in the large-$g$ region (due to the large $g_{0}$). For the plane-wave
phases (II) and (III), the latter has global ferromagnetic order $\langle
F_{z}\rangle \neq 0$ and is hence favorable with strong interaction. The $%
\Delta $-$g$ diagram also shows first-order transitions between any two of
(III), (IV), and (V) phases, second-order transitions between any other
adjacent phases, and four triple points $C_{1,2,3,4}^{\mathrm{T}}$ at the
(I)-(II)-(V), (II)-(III)-(V), (III)-(IV)-(V), and (I)-(IV)-(V) encounters,
respectively. In Fig.~\ref{fig:phase}(d), we show how the encounters of
phases along $\Delta =0$ change with the interaction. We see that phases
(III) and (IV) survive at large $g$, while (I) and (II) survive at large $%
\Omega $, in agreement with the energetic argument. The boundaries represent
three traces of triple points $C_{1,3}^{\mathrm{T}}$ and quadruple point $C_{%
\mathrm{quad}}$, respectively, which intercept at a quintuple point $C_{%
\mathrm{quin}}$ as the joint of all five phases.

In Figs.~\ref{fig:GP}(a) and (b), we plot spatial profiles of each spin
component's density $\rho _{\downarrow ,0,\uparrow }$ and total density $%
\rho _{t}$ for stripe phases (IV) and (V), respectively. Phase (IV) shows
out-of-phase modulations between $\rho _{\uparrow }$ and $\rho _{\downarrow
} $, representing spin-vector ($F_z$) density wave, and uniform $\rho _{0}$
and $\rho _{t}$, while (V) shows in-phase modulations of all components and
hence $\rho _{t}$, of which $\rho _{\uparrow ,\downarrow }$ overlap each
other, representing a spin-tensor ($F^2_{z}$) density wave. The modulation
wavelength matches the laser's recoil momentum $2k_{R} $ (i.e., $%
|k_{2}-k_{1}|=2k_{R}$). This can be understood in the quasi-momentum frame
that the minimization of $g_{2}$ interaction energy requires equal modulations
between the spin components and the spin operator $\mathbf{F}_{U}$ in Eq.~(%
\ref{eq:energydensity}). Since the separation between two band minima is
smaller than $2k_{R}$ at finite $\Omega $, the two plane-wave components of
the stripe phases do not exactly stay on the band minima. In Figs.~\ref%
{fig:GP}(c) and (d), we plot $\langle F_{z}\rangle $ (squares) and $\langle
F_{z}^{2}\rangle $ (circles) along (III)-(II) and (III)-(V)-(IV)-(I)
transition paths in Fig.~\ref{fig:phase}(b), respectively. The discontinuity
in spin-tensor polarization $\langle F_{z}^{2}\rangle $ (its first
derivative) indicates the occurrence of first-order (second-order) phase
transition.

\emph{Dynamical stripe state}.---The middle-band minimum and the right
bottom-band minimum are close to each other (both near $k=2$). Therefore a
coherent superposition of plane waves on these two minima leads to a
long-period stripe state, which can be directly measured in experiments. To
generate such a stripe state, we consider $^{\text{87}}$Rb atoms in a
harmonic trap $\omega =2\pi \times 50$Hz, initially prepared in spin state $%
\left\vert \uparrow \right\rangle $ with the Raman lasers off and $\Delta <0$
[the initial state belongs to phase (III) since the two minima coincide and
are equally populated as $\left\vert \uparrow \right\rangle =\frac{1}{\sqrt{2%
}}(\left\vert +\right\rangle +\left\vert -\right\rangle )$]. The 800-nm
Raman lasers are gradually turned on such that $\Omega $ increases from $0$
to $\Omega _{\text{f}}$ within a time $T$~and then remains constant. If we
consider an adiabatic process, where the ramping rate of $\Omega $ is much
slower than the energy scale of the spin-interaction strength $g_{2}\bar{n}$%
, the system will stay in the ground-state plane-wave phase (III) until $%
\Omega $ exceeds the critical value where a transition to plane-wave phase
(II) occurs. While for a dynamical process where the ramping rate of $\Omega
$ is much faster than the spin-interaction strength (but much slower than
other energy scales such as the trapping frequency), the system no longer
stays in the ground state, and the BEC on the two band minima are expected
to split in the momentum space, leading to the stripe state.

\begin{figure}[tbp]
\includegraphics[width=1.0\linewidth]{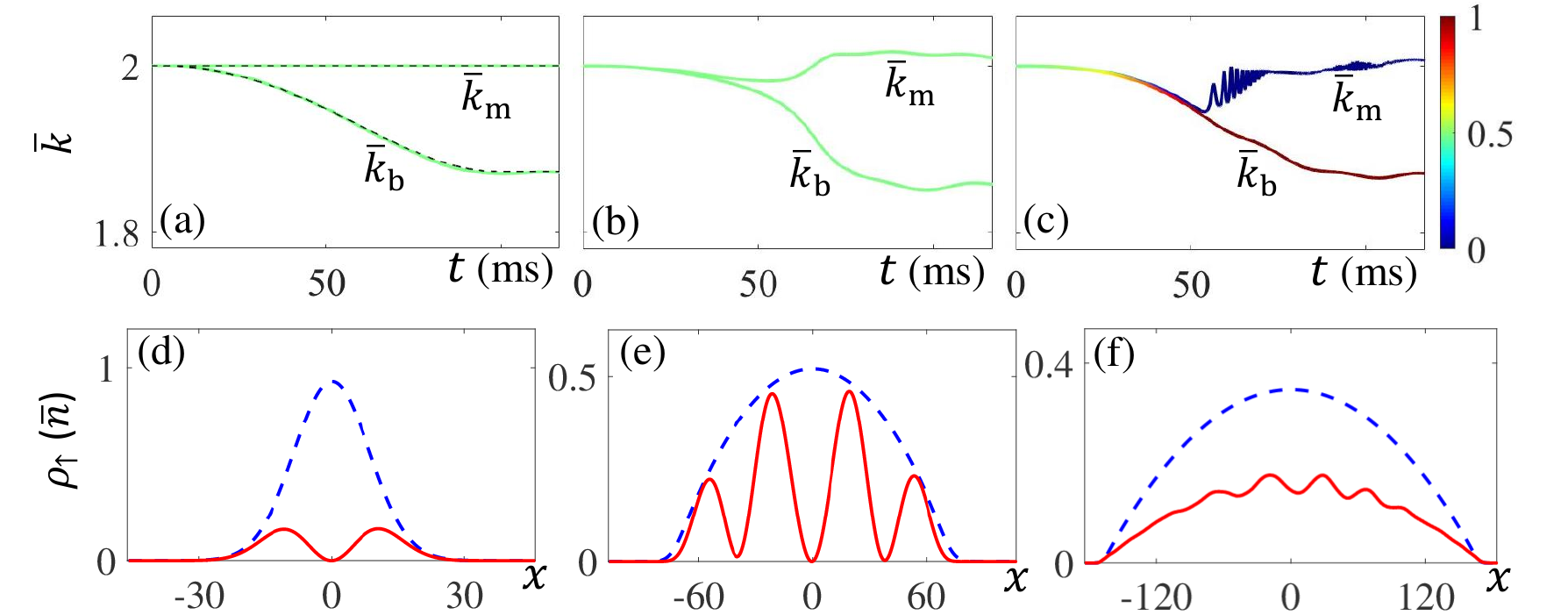}
\caption{ (a)-(c) Averaged momentum $\bar{k}_{m}$ ($\bar{k}_{b}$) and
percentage population (colorbar) of atoms in the middle (bottom) band. Thin
dashed line in (a) shows the band minima. (d)-(f) The initial (dashed line)
and final (solid line) spin density $\protect\rho _{\uparrow }$
corresponding to (a)-(c) respectively. The interactions are $(g_{0}\bar{n}%
,g_{2}\bar{n})=(0,0)$, $(0.5,0)$ and $(4,-0.2)$ for (a), (b) and (c). Other
parameters are $T=100$ms, $\Delta =-0.05$, and $\Omega_\text{f}=0.7$.}
\label{fig:Dyn}
\end{figure}

Figs.~\ref{fig:Dyn}(a) and (d) show the results of real-time GP simulation
for non-interacting atoms. The averaged momenta $\bar{k}_{\text{b}}$ and $%
\bar{k}_{\text{m}}$ of atoms in the bottom and middle bands follow their
band minima respectively, with $\bar{k}_{\text{b}}$ displaying slight dipole
oscillation \cite{chen2012collective} at $t>T$ due to the collective
excitations caused by the finite increasing rate of $\Omega $. The final
state is a stripe state similar to phase (IV) but with a much higher
visibility and a longer period, and the stripe pattern is moving rather than
stationary due to the dynamical phases of the two bands \cite{SM}.

For atoms with realistic interactions $|g_{2}|\ll g_{0}$ and consider a
dynamical process much faster compared to $g_{2}\bar{n}$, we can neglect the
spin interaction and focus on the density-interaction effects. The density
interaction preserves the symmetry $F_{x}^{2}$ and thus the atom populations
of the two bands remain unchanged. However, $\bar{k}_{\text{m}}$ shifts
together with $\bar{k}_{\text{b}}$ at the beginning then they separate and
eventually return to their band minima respectively. At $t>T$, the density
interaction induces synchronous dipole oscillations of $\bar{k}_{\text{m}}$
and $\bar{k}_{\text{b}}$ with a frequency different from the single-particle
case [see Fig.~\ref{fig:Dyn}(b)]. Nevertheless, we obtain a stripe state as
the final state [see Fig.~\ref{fig:Dyn}(e)] with a long period ($\sim 5\mu $%
m for $\Omega _{\text{f}}=0.7$) and high visibility (close to 100\%). For $^{%
\text{87}}$Rb with $g_{2}=-0.005g_{0}$, such dynamical stripe states can
always be obtained in the region where $|g_{2}|\bar{n}\ll T^{-1}$ \cite{SM}.
Also, the stripe period can be tuned by changing the value of $\Omega _{%
\text{f}}$ (e.g. $\Omega _{\text{f}}=1$ leads to a period of $\sim 3\mu $m)~%
\cite{SM}. Such periodic density modulations of dynamical stripe phases
break the translational symmetry of the Hamiltonian, showing dynamical
supersolid-like properties.

In the opposite region where the dynamical process is slow compared to the
spin interaction, the system follows the plane-wave ground state. As $\Omega
$ increases, atoms are transferred from the middle to bottom band until a
transition to phase (II) occurs. Thus the final state has no middle-band
population and no stripe states would be obtained, as shown in Figs.~\ref%
{fig:Dyn}(c) and (f) with tiny stripes caused by weak excitations.

\emph{Conclusions.}---In summary, we propose a scheme to realize STMC in a
spin-1 BEC, and study its ground-state and dynamical properties. The
interplay between STMC and atomic interactions leads to many interesting
quantum phases and multicritical points for phase transitions. The STMC
offers a simple way to generate a new type dynamical stripe states with high
visibility and long tunable periods, paving the way for direct experimental
observation of long-sought stripe states. The proposed STMC for ultracold
atoms open the door for exploring many other interesting physics, such as
STMC fermionic superfluids, Bogoliubov excitations with interesting roton
spectrum \cite{khamehchi2014measurement,Ji2015}, non-Abelian STMC (similar
as Rashba spin-orbit coupling), and STMC in optical lattices (where
nontrivial topological bands may emerge).

\begin{acknowledgments}
\textbf{Acknowledgements}: We thank P. Engels for helpful discussion. This
work is supported by AFOSR (FA9550-16-1-0387), NSF (PHY-1505496), and ARO
(W911NF-17-1-0128).
\end{acknowledgments}

%\bibliographystyle{apsrev}
%\bibliographystyle{unsrt}
%\bibliography{Spin1-Paper}

\newpage \newpage

\begin{widetext}

\section*{Supplementary Materials}
\setcounter{figure}{0}
\renewcommand{\thefigure}{S\arabic{figure}}
\setcounter{equation}{0}
\renewcommand{\theequation}{S\arabic{equation}}

\subsubsection*{Validation of the ansatz}
{The top and
bottom bright-state bands exhibit the same physics as the known
spin-orbit-coupled spin-$1/2$ system: the two spin branches $|0\rangle $ and
$|+\rangle $ with relative energy difference $\Delta $ are separated by $%
2k_{R}$ at $\Omega =0$, and mixed to form top/bottom bands with a gap at a
finite $\Omega $. At $\Delta =0$, the bottom band has degenerate double
minima for $\Omega <\sqrt{2}E_{\text{R}}$, above which the band makes a
transition to a single-minimum structure. The decoupling of the middle band
is protected by the spin-tensor symmetry $F_{x}^{2}$, under which
the middle band (top and bottom bands) corresponds to $\langle
F_{x}^{2}\rangle =0$ (1). Therefore, even if the gap between the middle and
bottom band minima is small [$\sim O(\Omega ^{2})$ at weak $\Omega $], the
single-particle ground state always selects one minimum on the the bottom band. However, the
atomic interactions can break the symmetry and drastically change the BEC's
ground state by involving the middle band.}
The ground state is mainly determined by the
two lower bands, with three minima in total. So we may consider a more general ansatz
\begin{eqnarray}
\Psi=\sqrt{\bar{n}}\left(|c_1|\chi_1e^{ik_1x} +|c_2|\chi_2e^{ik_2x+i\alpha}+|c_3|\chi_3e^{ik_3x+i\beta}\right),
\label{eq:ans-sm}
\end{eqnarray}
with $k_1\simeq 0$, $k_{2,3}\simeq 2$,
and $\chi_i=(\cos\theta_i\cos\phi_i,-\sin\theta_i,\cos\theta_i\sin\phi_i)^T$.
The stripe phase is supposed to lower the
spin interaction $g_2(\Psi^{\dag}\mathbf{F}_U\Psi)^2$ by generating ferromagnetic
order. The ferromagnetic order is maximized when
$k_{2}=k_3=k_1+2$, that is when the modulation of the spin density
is equal to the modulation of the spin operator $\mathbf{F}_U$. Then
Eq.~(\ref{eq:ans-sm}) is reduced to the
ansatz given in the main text.
The above arguments are verified
numerically by considering the ansatz Eq.~(\ref{eq:ans-sm})
and we always have $k_3=k_2=k_1+2$ for the ground state.

\subsubsection*{Variational energy density}
In the following, we give a detailed derivation
of the variational energy density, using the variational ansatz
\begin{eqnarray}
\Psi=\left(\begin{array}{c}
    \psi_{+1}  \\
    \psi_{0}  \\
    \psi_{-1}   \\
  \end{array}\right)=\sqrt{\bar{n}}|c_1|\left(
  \begin{array}{c}
    \cos(\theta_1)\cos(\phi_1)   \\
    -\sin(\theta_1)  \\
    \cos(\theta_1)\sin(\phi_1)   \\
  \end{array}
\right)e^{ik_1x}
+\sqrt{\bar{n}}|c_2|\left(
  \begin{array}{c}
    \cos(\theta_2)\cos(\phi_2)   \\
    -\sin(\theta_2)  \\
    \cos(\theta_2)\sin(\phi_2)   \\
  \end{array}
\right)e^{ik_2x+i\alpha}.
\end{eqnarray}
The single particle energy density is
\begin{eqnarray}
\varepsilon_0=\frac{1}{V}\int\Psi^\dag H_0\Psi dx=\frac{1}{V}\int\Psi^\dag[(-\partial^2_x) + (\Delta+4+4i\partial_x) F_z^2 + \sqrt{2}\Omega F_x]\Psi dx.
\end{eqnarray}
We have
\begin{equation}
\frac{1}{V}\int\Psi^\dag (-\partial^2_x)\Psi dx=\frac{1}{V}\int\sum_{j=0,\pm1}\psi^*_j (-\partial^2_x)\psi_j dx=\bar{n}\sum_{i=1}^2|c_i|^2k_i^2,
\end{equation}
similarly we can obtain
\begin{equation}
\frac{1}{V}\int\Psi^\dag \sqrt{2}\Omega F_x\Psi dx=-\bar{n}\sqrt{2}\Omega \sum_{i=1}^2|c_i|^2\sin(2\theta_i)\sin(\phi_i+\frac{\pi}{4}),
\end{equation}
and
\begin{equation}
\frac{1}{V}\int\Psi^\dag (\Delta+4+4i\partial_x) F_z^2\Psi dx=\bar{n}\sum_{i=1}^2(\Delta+4-4k_i)|c_i|^2\cos^2(\theta_i).
\end{equation}

The density-interaction energy is
\begin{eqnarray}
\varepsilon_{\text{d}}&=&\frac{1}{V}\int dx \frac{g_0}{2} (\Psi^{\dag} \Psi)^2=
\frac{g_0}{2}\frac{1}{V}\int dx  \left(\sum_{j=0,\pm1}|\psi_j|^2\right)^2 \nonumber\\
&=&\bar{n}\frac{g_0\bar{n}}{2}\left\{1+2|c_1|^2|c_2|^2[\sin(\theta_1)\sin(\theta_2)+\cos(\theta_1)\cos(\theta_2)\cos(\phi_1-\phi_2)]^2\right\},
\end{eqnarray}
and the spin-interaction energy is
\begin{eqnarray}
\varepsilon_\text{s}=\frac{1}{V}\int dx \frac{g_2}{2} (\Psi^{\dag}\mathbf{F}_U\Psi)^2,
\end{eqnarray}
with spatially modulated spin operator $\mathbf{F}_U=(F_U^x,F_U^y,F_U^z)$,
\begin{equation}
F_U^x=\frac{1}{\sqrt{2}}\left(
  \begin{array}{ccc}
    0 & e^{i2k_Rx} & 0  \\
    e^{-i2k_Rx} & 0 & e^{-i2k_Rx}  \\
    0 & e^{i2k_Rx} & 0  \\
  \end{array}
\right)
\end{equation}

\begin{equation}
F_U^y=\frac{1}{\sqrt{2}}\left(
  \begin{array}{ccc}
    0 & -ie^{i2k_Rx} & 0  \\
    ie^{-i2k_Rx} & 0 & -ie^{-i2k_Rx}  \\
    0 & ie^{i2k_Rx} & 0  \\
  \end{array}
\right)
\end{equation}
and
$F_U^z=F_z$.
Thus we have
\begin{eqnarray}
\varepsilon_\text{s}
&=&\frac{g_2}{2}\frac{1}{V}\int dx \left[\left(|\psi_{+1}|^2-|\psi_{-1}|^2\right)^2+2\left|\psi_0^*\psi_{+1}e^{-i2k_{\text{R}}x}+\psi_{-1}^*\psi_0e^{i2k_{\text{R}}x}\right|^2\right]\nonumber \\
&=&\bar{n}\frac{g_2\bar{n}}{2} \big\{2|c_1c_2|^2\cos^2(\theta_1)\cos^2(\theta_2)\cos^2(\phi_1+\phi_2) + |c_1c_2|^2\sin(2\theta_1)\sin(2\theta_2)\cos(\phi_1-\phi_2) \nonumber\\
& & +\left[\sum\nolimits_i|c_i|^2\cos^2(\theta_i)\cos(2\phi_i)\right]^2 +2\left[\sum\nolimits_i|c_i|^2\sin^2(\theta_i)\right]\left[\sum\nolimits_i|c_i|^2\cos^2(\theta_i)\right] \nonumber\\
& & +2\delta_{k_1,k_2-2}|c_1c_2|^2\sin^2(\theta_1)\cos^2(\theta_2)\sin(2\phi_2)\cos(2\alpha)\big\}.
\end{eqnarray}
Then we obtain the total energy density as
\begin{equation}
\varepsilon=\varepsilon_0+\varepsilon_\text{d}+\varepsilon_\text{s}.
\end{equation}
The stripe phase is supposed to lower the spin-interaction energy density $\varepsilon_\text{s}$,
in which there exists a term proportional to $\delta_{k_1,k_2-2}$.
This gives the mathematic reason why we always have
$k_2-k_1=2$ in the stripe phases.

{The variational ansatz leads to an energy
density which is a functional of  eight parameters.
Such an energy density plays the role of
Ginsburg-Landau potential, and the
ground state and the corresponding energy density are obtained by finding
the minimum of the Ginsburg-Landau potential with respect to all eight
parameters.}
{The quantum
phase diagram can be characterized with the variational wavefunction,
experimental observables $\langle F_{z}\rangle $ and $\langle
F_{z}^{2}\rangle $, and the symmetry property $\langle F_{x}^{2}\rangle $.
The phase transitions in our system are determined based on the Ehrenfest classification, with the
order of the phase transition labeled by the lowest derivative of the
ground-state energy density $\varepsilon_\text{g}=\min\{\varepsilon\} $ that is discontinuous at the
transition. In particular, we examine the
derivatives $\frac{\partial \varepsilon_\text{g} }{\partial \Delta }= \langle
F_{z}^{2}\rangle $ and $\frac{\partial ^{2}\varepsilon_\text{g} }{\partial \Delta ^{2}%
}= \frac{\partial \langle F_{z}^{2}\rangle }{\partial \Delta }$
(One can apply the Hellmann-Feynman theorem to obtain these relations),
and $\langle F_{z}^{2}\rangle $ ($\frac{\partial
\langle F_{z}^{2}\rangle }{\partial \Delta }$) displays discontinuity as $%
\Delta $ varies across a first-order (second-order) phase boundary  [see
Figs. 3(c) and (d) in the main text].
This argument also applies to the
derivatives $\frac{\partial \varepsilon_\text{g}}{\partial \Omega}$ ($\frac{\partial^2 \varepsilon_\text{g}}{\partial \Omega^2}$) as shown in Fig.~\ref{fig:sm0}, though they are less experimentally accessible. For a
crossing over, all these derivatives should be continuous.}

\begin{figure}[!htbp]
\includegraphics[width=0.5\linewidth]{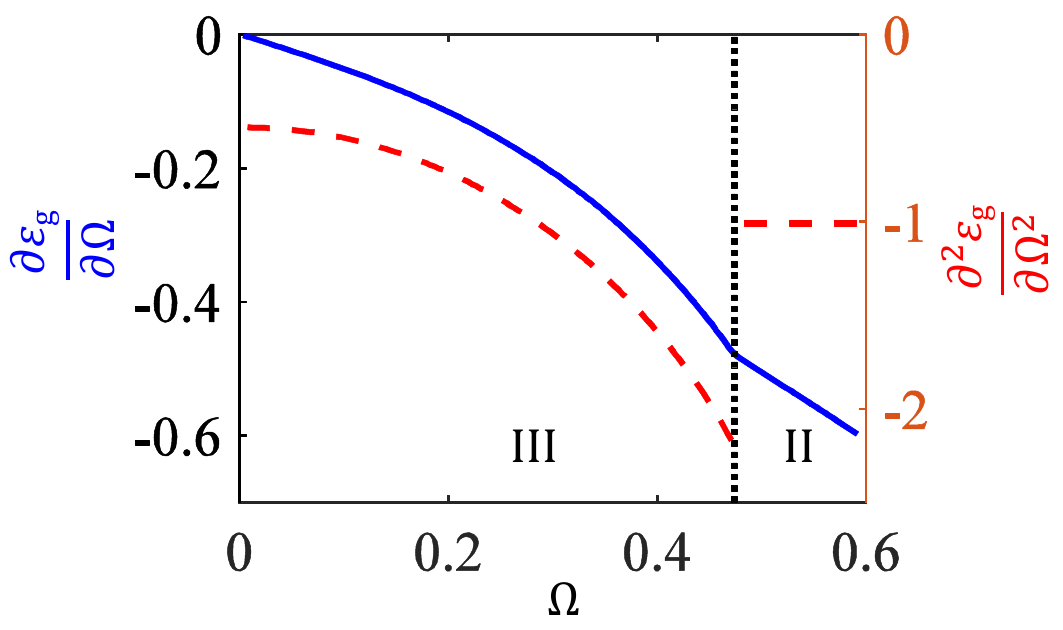}
\caption{ 
{First (blue solid line) and second (red dashed line) order derivatives of the ground state energy over
the Raman coupling strength $\Omega$. The discontinuity in the second order derivative implies
the transition between phases (II) and (III) is a second order one
(with the boundary given by the black-dotted vertical line).
Other parameters are the same as in Fig.~3(c) in the main text.}}
\label{fig:sm0}
\end{figure}

\subsubsection*{Perturbation analysis}
We consider the regime where $\Omega$ and $\Delta$ are small, and
the interactions are weak.
For the ground state properties,
we can omit the high-energy top band safely, and
consider only the two lower bands.
The middle band has a minimum at $k=2$ with spin state
\begin{eqnarray}
 \chi_{\text{m}}=\left(
  \begin{array}{ccc}
    \frac{1}{\sqrt{2}},   &
    0,  &
    \frac{-1}{\sqrt{2}}
  \end{array}
\right)^T.
\end{eqnarray}
The bottom band has two minima, one at $k\simeq0$
with spin state
\begin{eqnarray}
 \chi_{\text{b,l}}=\left(
  \begin{array}{ccc}
    -\frac{\Omega}{4},  &
    1-\frac{\Omega^2}{16},  &
    -\frac{\Omega}{4}
  \end{array}
\right)^T,
\end{eqnarray}
and the other at $k\simeq2$ with spin state
\begin{eqnarray}
 \chi_{\text{b,r}}=\left(
  \begin{array}{ccc}
    \frac{1-\frac{\Omega^2}{16}}{\sqrt{2}}, &
    -\frac{\Omega}{2\sqrt{2}}, &
    \frac{1-\frac{\Omega^2}{16}}{\sqrt{2}}
  \end{array}
\right)^T.
\end{eqnarray}

\begin{figure}[!htbp]
\includegraphics[width=0.75\linewidth]{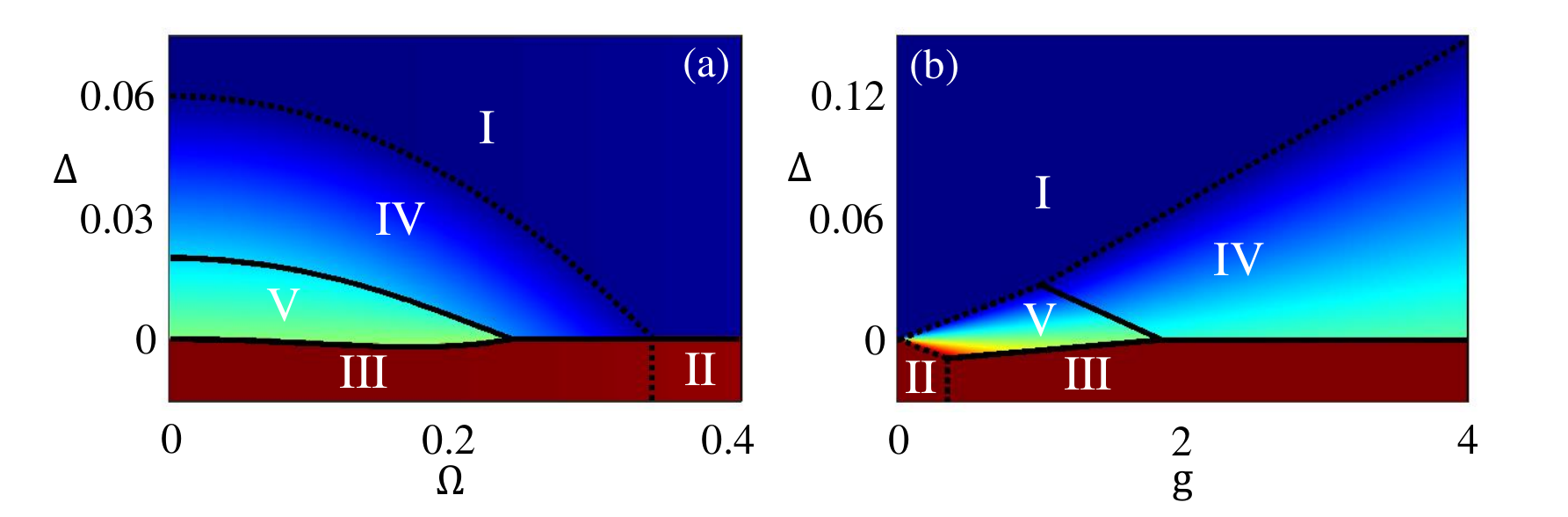}
\caption{ (a)  Phase diagram in the $\Omega$-$\Delta$ plane
with $g=1.5$. (b) Phase diagram in the $\Delta$-$g$ plane
with $\Omega=0.16$.
Solid lines represent first order
phase transitions while dotted lines represent second order phase transitions.
The phase diagram is obtained using perturbation analysis with interaction ratio $g_0=-50g_2\equiv g$.}
\label{fig:sm1}
\end{figure}

As we discussed above, the ground state may contain two plane waves at most,
so we consider a perturbation ansatz
\begin{eqnarray}
\Psi_\text{p}=|c_1|\chi_{\text{b,l}}e^{ik_1x} +\left(|c_2|\chi_{\text{b,r}}e^{i\alpha}
+|c_3|\chi_{\text{m}}e^{i\beta}\right)e^{i(k_1+2)x},
\end{eqnarray}
with $|c_1|^2+|c_2|^2+|c_3|^2=1$, the energy density now becomes
\begin{eqnarray}
\varepsilon&=&-|c_1|^2\frac{\Omega^2}{2}+|c_2|^2(\Delta-\frac{\Omega^2}{2})+|c_3|^2\Delta
+ \frac{g_0}{2}\left(1+|c_1|^2|c_2|^2\Omega^2\right) \nonumber\\
& &+g_2\left[|c_1|^2(|c_2|^2+|c_3|^2)+2|c_2c_3|^2\cos^2(\alpha-\beta)\right]
+g_2\left[|c_1c_2|^2\cos(2\alpha)-|c_1c_3|^2\cos(2\beta)\right]
\end{eqnarray}
According to the second partial derivative test,
it can be proven that the minima of $\varepsilon$ always satisfy $c_1c_2c_3=0$,
which means that the co-occupation of three band minima is never energetically favorable.
So there are three cases:

(1) $c_1=0$ and $\cos^2(\alpha-\beta)=1$,
$\Psi_\text{p}$ describes a plane-wave state in phase (II) or a polarized plane-wave state in phase
(III), with its energy density
\begin{eqnarray}
\varepsilon_{23}&\equiv&\varepsilon|_{c_1=0} =|c_2|^2(\Delta-\frac{\Omega^2}{2})+|c_3|^2\Delta
+ \frac{g_0}{2}+2g_2|c_2c_3|^2.
\label{eq:E23}
\end{eqnarray}

(2) $c_2=0$ and $\sin^2(\beta)=1$, $\Psi_\text{p}$ describes a plane-wave state in phase (I) or stripe state in phase (IV) with
energy density
\begin{eqnarray}
\varepsilon_{31}&\equiv&\varepsilon|_{c_2=0} =-|c_1|^2\frac{\Omega^2}{2}+|c_3|^2\Delta + \frac{g_0}{2}+2g_2|c_1c_3|^2.
\label{eq:E31}
\end{eqnarray}

(3) $c_3=0$ and $\cos^2(\alpha)=1$, $\Psi_\text{p}$ describes a plane-wave state in phase (I) or stripe state in phase (V) with
energy density
\begin{eqnarray}
\varepsilon_{12}&\equiv&\varepsilon|_{c_3=0} =-|c_1|^2\frac{\Omega^2}{2}+|c_2|^2(\Delta-\frac{\Omega^2}{2}) + \frac{g_0}{2}\left(1+|c_1|^2|c_2|^2\Omega^2\right) +2g_2|c_1c_2|^2.
\label{eq:E12}
\end{eqnarray}
{Generally, the Ginsburg-Landau potential can \emph{not} be written as a
functional of a single scalar order parameter for the interacting
multi-component bosonic fields considered here. Nevertheless, by assuming
an perturbative ansatz with fixed spin state and reduced parameter space,
the effective Ginsburg-Landau potential can be written as a
functional of a single scalar order parameter (either $c_1$ or $c_2$) for certain
phase transitions, as
can be seen from Eqs.~(\ref{eq:E23}), (\ref{eq:E31}, (\ref{eq:E12}).}

Therefore, the ground state is determined by minimizing $\varepsilon_{12},\varepsilon_{23},\varepsilon_{31}$,
with ground-state energy density given by
$\varepsilon_{\text{g}}=\min\{\varepsilon_{12},\varepsilon_{23},\varepsilon_{31}\}$.
Using Eqs.~(\ref{eq:E23}), (\ref{eq:E31}, (\ref{eq:E12}),
it is straight forward to calculate the ground state
and the corresponding energy density $\varepsilon_{\text{g}}$.
The phase boundaries
can be obtained by examining the ground state or
the derivation of $\varepsilon_{\text{g}}$ over
$\Omega, \Delta, \cdots$.
As shown in Figs.~\ref{fig:sm1} (a) (b), we find that
the phase boundary between (I) and (II) [(III) and (IV)] is
$\Delta=0$,
the phase boundary between (I) and (IV) is
$\Delta=-2g_2-\Omega^2/2$, the boundary between (II) and (III)
is $\Omega^2=-4g_2$, and the boundary between (V) and (I) [(II)] is $\pm\Delta=2g_2+g_0\Omega/2$.
The phase diagrams by perturbation analysis, as well as the behavior of the multicriticalities,
are qualitatively in good agreement with the full variational calculation, though
the exact phase boundaries are slightly different. This is because
the perturbation results are valid only to the order of $(g_2,\Delta,\Omega^2)$,
and generally the spin states of interacting BECs are slightly different from
the spin states in the perturbation ansatz.

\begin{figure}[!htbp]
\includegraphics[width=0.8\linewidth]{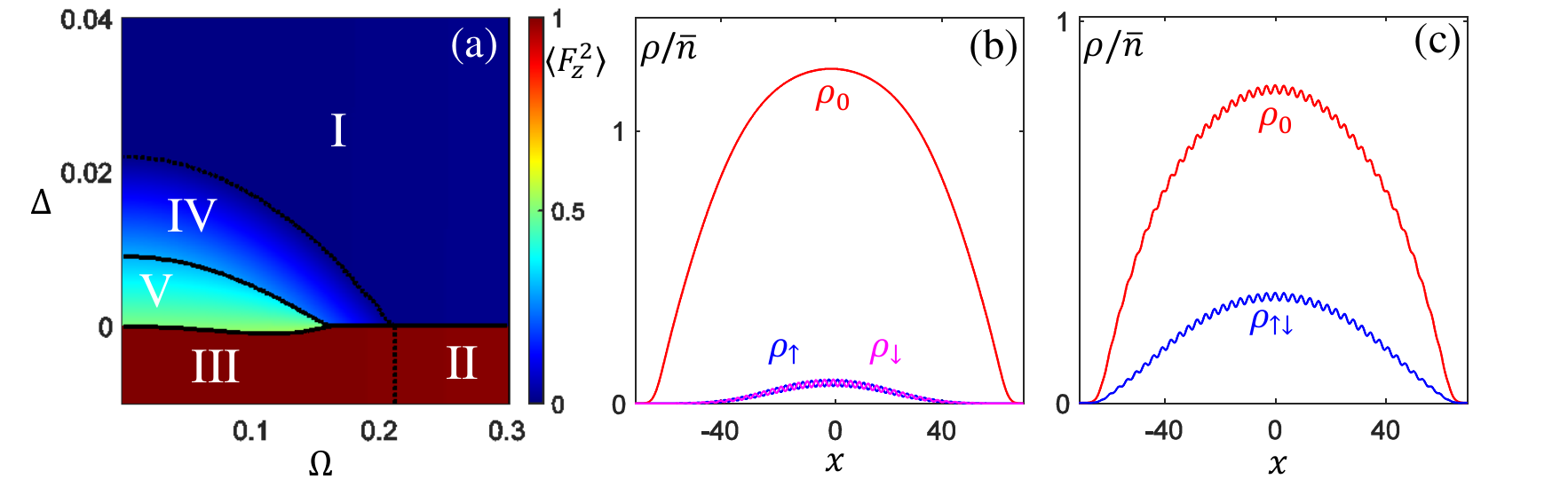}
\caption{ (a) Ground state phase diagram in the $\Omega$-$\Delta$ plane for
$^{87}$Rb BECs.
Solid lines represent first order
phase transitions while dotted lines represent second order phase transitions.
(b) Density modulations of stripe phase (IV) in the presence of harmonic trap, with $\Omega=0.08$, $\Delta=0.02$.
(c) Density modulations of stripe phase (IV) in the presence of harmonic trap, with $\Omega=0.08$, $\Delta=0.005$.
In (a-c), typical $^{87}$Rb interaction ratio $g_0=-200g_2$ is used, with $g_0\bar{n}=2.2$.}
\label{fig:sm2}
\end{figure}

\subsubsection*{Effects of interaction ratio and harmonic trap}
In the main text, we have fixed the interaction ratio as $g_0=-50g_2$,
a stronger (weaker) $g_2$ will enlarge (shrink) the regions
of stripe and polarized plane-wave phases,
but does not qualitatively change the phase diagram structure.
To show this, in Fig.~\ref{fig:sm2}(a), we give the phase diagrams of
interaction ratio $g_0=-200g_2$ for
$^\text{87}$Rb atoms.
Typically, the atomic density is about
$10^{15}$cm$^{-3}$, for s-wave scattering length 100.48$a_0$
($a_0$ is the Bohr radius) and Raman-laser wavelength $800$nm,
the corresponding interaction is
$g_0\bar{n} = 2.2$.
Moreover, for realistic experiments, the
BECs are confined by
a harmonic trap, we consider a trapping frequency
$\omega=2 \pi\times 0.2$kHz and calculate the ground state
using imaginary time evolution of the GP equation.
Figs.~\ref{fig:sm2} (b) and (c) show the ground-state density modulations
corresponding to stripe phases (IV) and (V).

\begin{figure}[!htbp]
\includegraphics[width=0.75\linewidth]{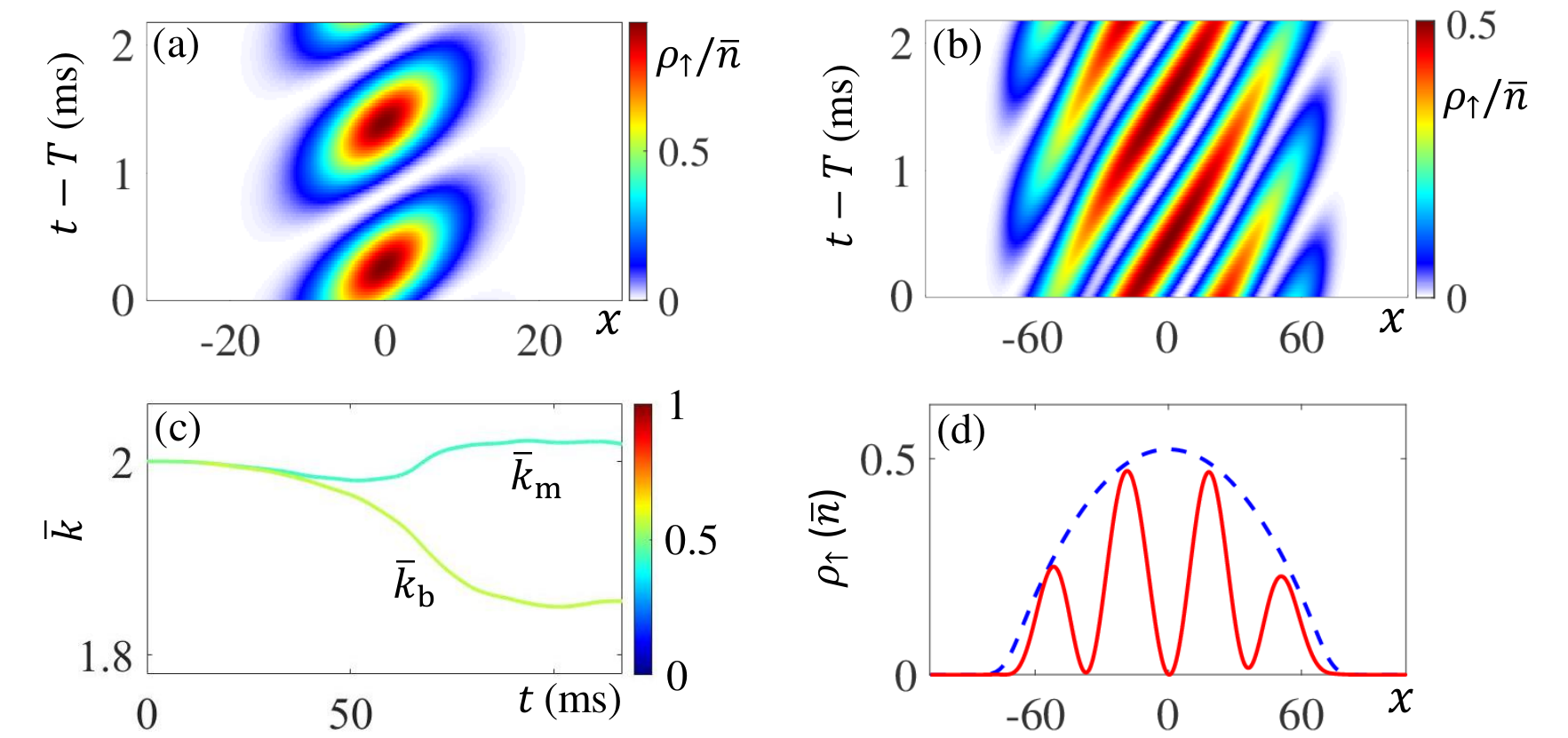}
\caption{ (a) and (b) The evolution of the spin density
corresponding to Figs. 4(d) and (e) in the main text.
(c) and (d) The same as in Figs. 4(b) and (e) in the main text
except that the interaction ratio of $^{87}$Rb ($g_0=-200g_2$) is used,
with $g_0\bar{n}=0.5$.}
\label{fig:sm3}
\end{figure}

\begin{figure}[!htbp]
\includegraphics[width=0.8\linewidth]{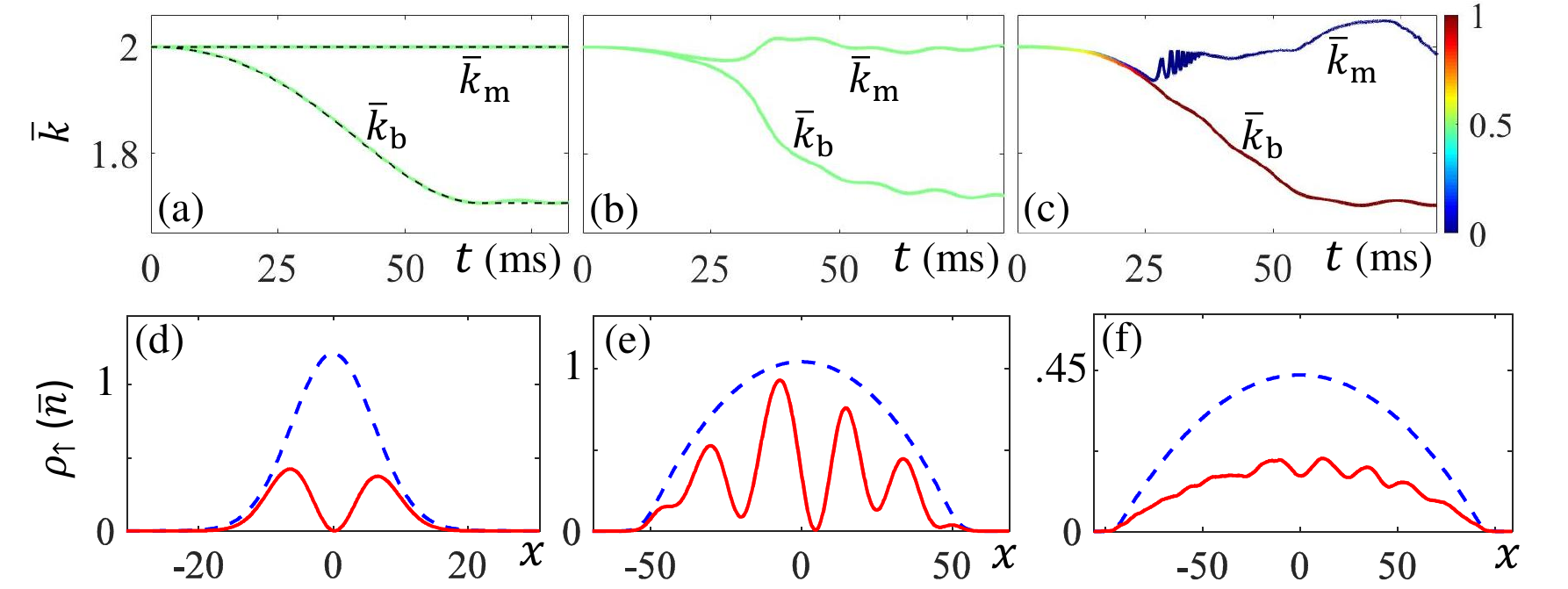}
\caption{ (a)-(c) Averaged momentum $\bar{k}_{m}$ ($\bar{k}_{b}$) and
percentage population (colorbar) of atoms in the middle (bottom) band. Thin
dashed line in (a) shows the band minima. (d)-(f) The initial (dashed line)
and final (solid line) spin density $\protect\rho _{\uparrow }$
corresponding to (a)-(c) respectively. The interactions are $(g_{0}\bar{n},g_{2}\bar{n}%
)=(0,0)$, $(0.5,0)$ and $(4,-0.2)$ for (a), (b) and (c).
The final value of $\Omega$ is $\Omega_\text{f}=1$, leading to a stripe period of $\sim3\mu$m.
$\omega=2\pi\times100$Hz is used to reduce the time period with
$T=65$ms. The detuning is $\Delta =-0.05$.}
\label{fig:sm4}
\end{figure}

\subsubsection*{Dynamical stripe states}
{Our dynamical process (where the ramping rate of $\Omega$ is
much faster than the spin-interaction strength) leads to a final state being
a nearly equal superposition of two plane waves (with an overall Gaussian-packet form
in the presence of a Harmonic trap),
\begin{equation}
\Psi =c_{\text{b}}\chi _{\text{b}}e^{ik_{\text{b}}x}+c_{\text{m}}\chi _{%
\text{m}}e^{ik_{\text{m}}x+i\phi _{\text{m}}(t)},
\end{equation}%
where \textquotedblleft b" (\textquotedblleft m") labels the bottom (middle)
band, with spin states $\chi _{\text{b(m)}}$, momentum $k_{\text{b(m)}}$ and
coefficients $c_{\text{b(m)}}\simeq \frac{1}{\sqrt{2}}$.
Although the equal superposition remains over time, $\Psi $ is
different from the ordinary stripe state by a dynamical phase $\phi _{\text{m%
}}(t)$ originating from the energy difference between middle and bottom
bands.}

{Nevertheless, this state has a uniform total density and a striped
sinusoidal spin density. Although the spin-density modulation propagates in
space due to the dynamical phase $\phi _{\text{m}}(t)$, the visibility and
period of the spin-density modulation do not change, as shown in Figs.~\ref{fig:sm3} (a) and (b).
Furthermore, the
dynamical stripe state itself has the superfluid property and breaks the
translational symmetry of the Hamiltonian, showing a supersolid-like
property. Note that
the density modulation period is tunable and long enough for direct
experimental observation of such dynamical stripe state.}

%For the dynamical stripe states, the stripe pattern is moving rather than stationary
%due to the dynamical phases of the two bands, as shown in Figs.~\ref{fig:sm3} (a) and (b).
The long-period and high-visibility dynamical stripe states
can always be obtained as long as the spin interaction is weak,
as shown in Figs.~\ref{fig:sm3} (c) and (d), where we consider
the weakly interacting $^\text{87}$Rb atoms with $g_0\bar{n}=0.5$ and typical interaction ratio
$g_0=-200g_2$.
The population percentage of the
bottom band is slightly increased since some middle-band atoms are scattered
to the bottom band by spin interaction, and the stripe pattern in Fig.~\ref{fig:sm3} (d) moves similarly as in Fig.~\ref{fig:sm3} (b).
Moreover, the modulation period can be tuned by changing the value of $\Omega_{\text{f}}$,
as shown in Fig.~\ref{fig:sm4} with $\Omega_{\text{f}}=1$ and
a corresponding stripe period of $\sim3\mu$m.
In Fig.~\ref{fig:sm4} (e), the stripe visibility is slightly reduced,
because the spin component $\vert0\rangle$ in the bottom band increases slightly
with $\Omega_\text{f}$.

\end{widetext}

\end{document}